\documentclass[11pt]{article}

\def\laq{~\raise 0.4ex\hbox{$<$}\kern -0.8em\lower 0.62
ex\hbox{$\sim$}~}
\def\gaq{~\raise 0.4ex\hbox{$>$}\kern -0.7em\lower 0.62
ex\hbox{$\sim$}~}

\def\beq{\begin{equation}}
\def\eeq{\end{equation}}
\def\bea{\begin{eqnarray}}
\def\eea{\end{eqnarray}}
\newcommand{\nn}{\nonumber}
\newcommand{\de}{\partial}

\def \tpsi {\widehat \psi}
\def \Mp {M_{\rm P}}

\def \pa {\partial}

\def \ra {\rightarrow}

\def \La {\Lambda}

\def \b {\beta}
\def \a {\alpha}

\def \Ga {\Gamma}
\def \ga {\gamma}

\def \da {\delta}
\def \ep {\epsilon}

\begin{document}

\begin{titlepage}

\begin{flushright}
BA-TH/04-492\\
hep-th/0501251
\end{flushright}

\vspace*{1.5 cm}

\begin{center}

\Huge
{Relic gravitons on Kasner-like branes}

\vspace{1cm}

\large{M. Cavagli\`a$^1$, G. De Risi$^{2,3}$ and M. Gasperini$^{2,3}$}

\normalsize
\vspace{.2in}

{\sl $^1$Department of Physics and Astronomy, University of Mississippi,\\ 
MS 38677-1848, USA}\\

\vspace{.2in}

{\sl $^2$Dipartimento di Fisica , Universit\`a di Bari, \\ 
Via G. Amendola 173, 70126 Bari, Italy}

\vspace{.2in}

{\sl $^3$Istituto Nazionale di Fisica Nucleare, Sezione di Bari\\
Via G. Amendola 173, 70126 Bari, Italy}

\vspace*{1.5cm}

\begin{abstract}
We discuss the cosmological amplification of tensor perturbations in a simple
example of brane-world scenario, in which massless gravitons are localized on a
higher-dimensional Kasner-like brane embedded in a bulk AdS background. 
Particular attention is paid to the canonical normalization of the quadratic
action describing the massless and massive vacuum quantum fluctuations, and
to the exact mass-dependence of the amplitude  of massive fluctuations on the
brane. The perturbation equations can be separated. In contrast to de
Sitter models of brane inflation, we find no mass gap in the spectrum and no
enhancement for massless modes at high curvature. The massive modes can
be amplified, with  mass-dependent amplitudes, even during inflation and in the
absence of any mode-mixing effect. 
\end{abstract}
\end{center}

\end{titlepage}
\newpage

\parskip 0.2cm

The amplification of the metric quantum fluctuations induced by inflation is
expected to produce a cosmic background of relic gravitons,  carrying unique
information on the primordial state of our universe \cite{1}. This effect is
also expected to occur in ``brane-world cosmology" models  (see
e.g.~Ref.~\cite{2}.) Brane-world cosmology was originally inspired by heterotic
M-theory \cite{3}, and is phenomenologically motivated by the Randall-Sundrum
mechanism \cite{4}, which allows the localization of long-range gravitational
interactions on the brane through the warping of the higher-dimensional bulk. 

Recent studies on inflationary graviton production in the brane-world scenario
gave interesting results \cite{5,6,6b} on the possible mixing of massless and massive
modes of the tensor perturbation spectrum, and on the possible enhancement of the
massless spectral amplitudes occurring above a threshold value of the brane
curvature. In these models, the geometry of the brane can be parametrized as a
four-dimensional de Sitter slicing of a five-dimensional anti-de Sitter (AdS)
manifold. The aim of this letter is to point out that some of the results are
specific to the models of inflation considered and cannot be extended to
cosmological graviton production in a generic brane-world scenario. 

In this letter we consider a model based on a higher-dimensional Kasner-like
brane embedded in an AdS bulk manifold. There are no matter sources on the
brane; inflation (i.e. accelerated expansion) is sustained by the accelerated contraction  (``spontaneous"
dimensional reduction) of the brane internal dimensions.  Although the curvature of the brane is time-dependent and the geometry is not de Sitter, we find that the tensor perturbation equation separates in the brane and bulk coordinates. In contrast to the results of \cite{5,6}, the amplification
of the massless modes is always controlled by the brane curvature in the usual way, without anomalous
enhancement at higher curvature. Moreover, there is no mass gap in the spectrum. However, massive modes with mass smaller than the brane
curvature scale are also directly amplified during the phase of brane
inflation, and may lead to a significant enhancement of the final amplitude of the tensor perturbation background. 

Let us consider a $(p+2)$-dimensional configuration in which a $D_p$-brane is
embedded in a $(p+2)$-dimensional bulk manifold with one extra spatial
dimension and a negative cosmological constant. We denote by $D=p+2$ the total
number of dimensions and assume the $p$-dimensional brane space to be the
product of two maximally symmetric manifolds with $d$ and $n$ dimensions
($p=d+n$). The dynamics of this system is described by the action 
\bea
S~=~S_{\rm bulk} +
S_{\rm brane}&=& - {1\over 2}M^{(D-2)} \int d^{D} x\sqrt{\left|g \right|}
\left(R+2\Lambda \right)\nn \\ 
&& 
-\frac{T_p}{2} \int d^{p+1}\xi
\sqrt{\left| \gamma \right|} \left[\gamma^{\alpha \beta}
\de_\alpha X^A  \de_\beta X^B g_{AB}(X) - (p-1) \right]\,,
\label{1}
\eea
where $\Lambda$ is the negative bulk cosmological constant, $T_p$ is the
tension of the brane, and $M$ is a mass parameter that characterizes the
strength of bulk gravitational interactions (when $p=1$ one recovers the string-gravity self-sustained configuration discussed in \cite{7a}). 
Here $\de_\alpha X^A$ is a compact
notation for $\de X^A(\xi)/\de \xi^\alpha$, where $\xi^\a$ are the coordinates
spanning the $(p+1)$-dimensional world-volume of the brane, and $X^A=X^A(\xi)$
are functions describing the parametric embedding of the brane into the
bulk manifold. $\ga_{\a\b}(\xi)$ is an auxiliary field representing
the metric tensor on the brane. Capital Latin indices run from $0$ to
$D-1$, Greek indices from  $0$ to $p$. For the bulk coordinates we use the notation $x^A \equiv
\left(t,x^i,y^a,z\right)$, where $i,j$ and $a,b$ run from $1$ to $d$ and from $d+1$
to $d+n$, respectively. 

The variation of Eq.~(\ref{1}) w.r.t.\ $g$, $X$ and $\ga$ leads
to the Einstein equations
\beq
G_{A}^{~B} = \Lambda \da_A^B + \frac{T_p}{M^{(D-2)}}  \frac{1}{\sqrt{\left|g
\right|}} g_{AC}
 \int d^{p+1} \xi  \sqrt{\left| \gamma \right|}
\gamma^{\alpha \beta} \de_\alpha X^C  \de_\beta X^B 
\delta^D\left(x-X(\xi) \right)\,,  
\label{2} 
\eeq
the equation for the brane evolution
\beq 
\de_\alpha \left( \sqrt{\left| \gamma
\right|}\gamma^{\alpha \beta} \de_\beta X^B g_{AB}(X) \right) =
\frac{1}{2} \sqrt{\left| \gamma \right|}\gamma^{\alpha \beta}
\de_\alpha X^B  \de_\beta X^C \pa_A g_{BC}(X)\,, 
\label{3} 
\eeq
and the equation for the induced metric on the brane
\beq
\gamma_{\alpha \beta} = \de_\alpha X^A  \de_\beta X^B 
g_{AB}(X)\,, 
\label{4} 
\eeq
respectively. We look for solutions describing a $Z_2$-symmetric bulk geometry
and a locally anisotropic brane with spatial geometry given by the product of
two conformally flat manifolds with $d$ and $n$ dimensions, respectively.
Assuming that the brane is rigidly located at $z=0$, we set
\beq
g_{AB} = f^2(z)~ {\rm diag} \left(1,-a^2(t) \da_{ij},-b^2(t) \da_{ab},-1
\right),  ~~~~~~~
X^A = \delta^A_\alpha \xi^\alpha\,. 
\label{ansatz}
\eeq
The equation for the brane is trivially satisfied. The induced metric is
\beq
\gamma_{\alpha \beta}(\xi)=g_{\alpha \beta}(\xi)|_{z=0}. 
\label{indmetric}
\eeq
For this background, the $(00)$, $(ii)$, $(aa)$, and $(p+1,p+1)$
components of the Einstein equations are 
\bea
&&
-p F' -{1\over 2}p(p-1)F^2 +
\frac{1}{2}d(d-1)H^2+\frac{1}{2}n(n-1)G^2+ dn HG 
\nn\\
&& 
= \Lambda f^2 + \frac{T_p}{M^{(D-2)}}f \delta(z)\,, 
\label{7} \\
&&
p F' +{1\over 2}p(p-1)F^2  - (d-1) 
\dot H -\frac{1}{2}d(d-1)H^2 -n \dot G- \frac{1}{2}n(n+1)G^2 
 \nn\\
&& 
-(d-1)nHG=-\Lambda f^2 - \frac{T_p}{M^{(D-2)}} f \delta(z)\,, 
\label{8} \\
&&
p F' +{1\over 2}p(p-1)F^2 - d \dot H -
\frac{1}{2}d(d+1)H^2 -(n-1) \dot G 
- \frac{1}{2}n(n-1)G^2  \nn\\
&&
-d(n-1)H G=-\Lambda f^2 - 
\frac{T_p}{M^{(D-2)}} f \delta(z)\,, \label{9} \\
&&
{1\over 2}p(p+1)F^2 -
d\left(\dot H + \frac{d+1}{2}H^2 \right) -
n\left(\dot G + \frac{n+1}{2}G^2 \right)
 -dn HG = - \Lambda f^2 \,.
\label{10}
\eea
Dots (primes) denote differentiation w.r.t.\ $t$ ($z$), and $H=
\dot a/a$, $G= \dot b /b$, $F=f'/f$. 

The time- and $z$-dependent parts of the above equations can be separated. The
flat Minkowski brane ($H=G=0$) is a trivial solution. Looking for non-trivial,
power-law solutions we set 
\beq
a(t) = \left({t}/{t_0}\right)^{\alpha}\,, ~~~~~~~~~~~
b(t) = \left({t}/{t_0}\right)^{\b}\,. 
\label{11}
\eeq
The Ricci-flat Kasner solution is 
\beq
\alpha = \frac{1 \pm  \sqrt{{n}(d+n-1)/d}}{d+n}\,, ~~~~~~~~~~~
\b = \frac{1 \mp  \sqrt{{d}(d+n-1)/n}}{d+n}\,, 
\label{12}
\eeq
where $d \a^2 +n \b^2= d\a+n \b=1$. A particular configuration with $d=3$ inflationary
expanding dimensions can be obtained for $\a<0$ and a negative range of the
cosmic time parameter $t$, corresponding to the lower-sign branch of
Eq.~(\ref{12}) with $n>1$ contracting dimensions. Finally, from the $z$-dependent part
of the above equations, we obtain the ($Z_2$-symmetric) warp factor for the AdS bulk geometry, 
\beq
f(z) = \left(1+{|z|}/{z_0}\right)^{-1}\,, ~~~~~~~
z_0 = \sqrt{-{p(p+1)}/{2\Lambda}}\,, 
\label{13}
\eeq
where $\La$, $M$ and $T_p$ are related by
\beq
{2T_p}/{M^{(D-2)}}= \sqrt{-{32p\Lambda}/(p+1)}\,. 
\label{14}
\eeq
(See also \cite{8a} for warped Kasner-like solutions on a brane.) 

In order to discuss the propagation of metric perturbations on the above background we now introduce the expansion
\beq
g_{AB} \ra g_{AB} + \da g_{AB}\,, ~~~~~
\da g_{AB} \equiv h_{AB}\,, ~~~~~ \da g^{AB} =- h^{AB}\,, 
\label{15}
\eeq
and perturb equations (\ref{2})-(\ref{4})  to first order in $h_{AB}$. The
position of the brane is kept fixed by requiring $\da X^A=0$ (see for instance \cite{7}). In the linear approximation, different components of the tensor fluctuations $h_{AB}$ are decoupled. We are interested, in particular, in the transverse and traceless
perturbations $h_{ij}$ of the expanding $d$-dimensional part of the brane
metric (\ref{ansatz}) which, for $d=3$, describes the early-time geometry
of our present large-scale spacetime. Keeping $d$ generic, and choosing the
syncronous gauge, we thus set
\beq
h_{0A}=0, ~~~~~~ h_{aA}=0, ~~~~~~  h_{zA}=0, ~~~~~~ 
 g^{ij} h_{ij}=0= \nabla _j h_i^j\equiv \pa_j h_i^j\,.
\label{16}
\eeq
The indices of $h$ are raised and lowered with the unperturbed metric, and the covariant derivatives acting on $h$ are constructed with the unperturbed connection.

Assuming that the translations along the $n$ internal dimensions of the brane
are isometries not only of $g_{AB}$, but also of the full perturbed metric $g_{AB} = g_{AB} + \da g_{AB}$, we
can set 
\beq
h_{ij}=h_{ij} (t, x^i,z)\,.
\label{17}
\eeq
The perturbation of Eq.~(\ref{4}) for the induced metric gives 
\beq
\da \ga_{\mu\nu}= \left(\da_\mu^i \da_\nu^j h_{ij}\right)_{z=0}\,.
\label{18}
\eeq
The perturbation of Eq.~(\ref{3}) is identically satisfied. The perturbations
of the Einstein equations reduce to $\da R_A^B=0$. The propagation equation for
tensor metric fluctuations on the expanding part of the brane is 
\beq
\ddot h_i^j +\left(dH +n G\right) \dot h_i^j -{\nabla^2 \over a^2}  {h_i^j} -
{h_i^{\prime \prime j}} -p F  {h_i^{\prime j}}=0\,,
\label{19}
\eeq
where $\nabla^2=\da^{ij}\pa_i\pa_j$ is the flat-space Laplacian. As expected, each physical
polarization mode $ {h_i^j}$ satisfies the full covariant d'Alembert equation
$\nabla_A \nabla^A  {h_i^j}=0$, describing the free propagation of a
scalar mode on the brane. 

The correct normalization of $ {h_i^j}$ to an initial spectrum of  vacuum quantum fluctuations is obtained from the quadratic action corresponding to the equations of
motion (\ref{19}). This action is computed by perturbing Eq.~(\ref{1}) up to terms quadratic in the first-order fluctuations $\da g_{AB}$ and $\da \ga_{\mu\nu}$,  at fixed brane position $\da X^A=0$. We expand  the contravariant
components of the metric and the volume density to order $h^2$: 
\bea
&&
\da^{(1)} g^{AB}=-h^{AB}\,, ~~~~~~~~~~~~~~
\da^{(2)} g^{AB}=h^{AC}h_C^B\,, \nn \\
&&
\da^{(1)} \sqrt{|g|}=0\,,  ~~~~~~~~~~~~~~~~~~~
\da^{(2)} \sqrt{|g|}= -{1\over 4}  \sqrt{|g|} h_{AB} h^{AB}\,, 
\label{20}
\eea
and similarly the connection and the Ricci tensor. 
The second-order action takes contributions from $\sqrt{|g|}$, $R$ and $\ga$:
\bea
&&
\da^{(2)} S=
-{M^p\over 2} \int d^Dx \Bigg[\da^{(2)} \sqrt{|g|}\left(R+2\La\right)+
\nn \\ \qquad
&&
+\sqrt{|g|}\left(\da^{(2)} g^{AB} R_{AB}+ \da^{(1)} g^{AB}\da^{(1)}
R_{AB}+g^{AB}\da^{(2)} R_{AB}\right)\Bigg] \nn\\
&&  
-{T_p\over 2} \int d^{p+1}\xi \Bigg[\da^{(2)} \sqrt{|\ga|}\left(\ga^{\a\b}
\pa_\a X^A \pa_\b X^B g_{AB} -(p-1)\right) 
\nn \\
\qquad
&&+\sqrt{|\ga|}
\left(\da^{(2)} \ga^{\a\b}g_{AB}+\da^{(1)} \ga^{\a\b}\da^{(1)}
g_{AB}\right)\pa_\a X^A \pa_\b X^B \Bigg]\,.
\label{21}
\eea
The action for the perturbations of the auxiliary field $\ga_{\mu\nu}$ leads to
the constraint (\ref{18}). Using the latter, and the unperturbed equations
of motion, we obtain 
\beq
\da^{(2)} S ={M^p \over 8} \int d^Dx a^d b^n f^p \left[ 
\dot h_i^{~j} \dot h_j^{~i}-h_i^{\prime j} h_j^{\prime i}+ h_i^{~j}
{\nabla^2 \over a^2} h_j^{~i}\right]\,, 
\label{22}
\eeq
where we have integrated by parts. We can easily check that Eq.  (\ref{19}) is given by the
variation of Eq.~(\ref{22}) w.r.t.\ $h$. We set $h_{ij} =h_{(a)}
\ep^{(a)}_{ij}$, where $ \ep^{(a)}_{ij}$ is the spin-two polarization tensor, 
and the sum is over all the $(d+1)(d-2)/2$ independent polarization states.
Using Tr$[\ep^{(a)}\ep^{(b)}]=2 \da^{ab}$, Eq.~(\ref{22}) can be rewritten as
the sum over single-mode actions
\beq
\da^{(2)} S = \sum_{(a)} \da^{(2)} S_{(a)}\,, ~~~~~~~
\da^{(2)} S_{(a)} ={M^p \over 4} \int d^Dx a^d b^n f^p \left[ 
\dot h^2 -h^{\prime 2}+ h {\nabla^2 \over a^2} h\right]\,,
\label{23}
\eeq
where the polarization index $(a)$ on the scalar mode $h$ has been omitted in the last equation, for simplicity. 

The propagation equation (\ref{19}) can be separated in the bulk and brane
coordinates by setting
\beq
h(t,x^i,z)= \sum_m h_m(t,x^i,z)= \sum_m v_m(t, x^i) \psi_m(z)\,,
\label{24}
\eeq 
where the new variables $v, \psi$ are labelled by the mass eigenvalue $m$. (The
sum is replaced by integration over $m$ for the continuous part of the
eigenvalue spectrum.) We then obtain the equations 
\bea
&&
\ddot v_m +\left( d H +nG \right) \dot v_m -{\nabla^2\over a^2} v_m
=- m^2 v_m\,,\label{25} \\
&&
\psi_m^{\prime \prime} + p F \psi_m^\prime \equiv
f^{-p} \left(f^p \psi'\right)' = - m^2 \psi_m\,. 
\label{26}
\eea
In particular, the rescaled variable $\tpsi_m= \sqrt M f^{p/2} \psi_m$ satisfies the
Schrodinger-like equation 
\bea
&&
\tpsi_m^{\prime \prime}+ \left[m^2 -V(z)\right] \tpsi_m=0\,,
\nn \\
&&
V(z)= -{p\over z_0} \da(z) +{1\over 4} p (p+2) \left(z_0+|z|\right)^{-2}\,, 
\label{27}
\eea
whose effective potential $V(z)$ has a ``volcano-like" shape for $z_0>0$. As in
the standard Randall-Sundrum scenario \cite{4}, this  potential exactly
localizes the massless mode $m=0$ on the brane.  The solutions of Eq.
(\ref{26}) are normalized by requiring the variables $\tpsi_m$ to be
orthonormal w.r.t.\ inner products with measure $dz$, as in
conventional one-dimensional quantum mechanics \cite{4,5,6b,7,9a}. This is
equivalent to require the  $\psi_m$ to be orthonormal with measure $M dz f^p$ 
\cite{9b}:
\beq
\int dz M f^p \psi_m\psi_{m'}= \da(m,m'),
\label{28}
\eeq
where $\da(m,m')$ denotes a Kronecker symbol for the discrete part of the spectrum and a Dirac distribution for the continuous one. 

Inserting  the expansion (\ref{24}) in Eq. (\ref{23}), and using the orthonormality condition (\ref{28}) to integrate over $z$, the effective action can be written, modulo a total derivative, as a sum over the contributions of all fluctuation modes evaluated on the brane, 
$ \overline h_m \equiv h_m(t,x^i,0)=v_m (t,x^i) \psi_m(0)$: 
\beq
\da^{(2)} S= \sum_m 
\da^{(2)} S_m  = \sum_m {M^{d-1}\over 4 |\psi_m(0)|^2}
 \int d^{d+1}x~ a^{d} b^n 
\left(\dot{\overline h}_m^{2} +
\overline h_m {\nabla^2\over a^2} \overline h_m- m^2  \overline h_m^2\right)\,.
\label{29}
\eeq
We have used Eq. (\ref{26}) to eliminate $\psi_m^{\prime 2}$, and  have
omitted  the constant dimensionless volume factor $M^n \int d^ny$. (The
internal brane sections are assumed to be compactified on a comoving length
scale of order $M^{-1}$.) The above action can be written in canonical form by
introducing the conformal-time coordinate $\eta= \int dt/a$ and the canonical
variable $u_m$, 
\beq
u_m (\eta, x^i)= \xi_m(\eta) \overline h_m (\eta, x_i), ~~~~~~~~~~~
\xi_m(\eta)= \left(M\over 2 \right)^{d-1\over 2} {a^{(d-1)\over 2} b^{n\over 2}\over \psi_m(0)}.
\label{30}
\eeq
Integrating by parts, we find:
\beq
\da^{(2)} S =\sum_m{1 \over 2} \int d\eta ~d^dx \left(u_m^{\prime 2} +
u_m {\nabla^2} u_m-m^2 a^2 u_m^2 + {\xi_m^{\prime \prime}\over \xi_m} u_m^2
\right)\,,
\label{31}
\eeq
where, from now on, a prime will denote differentiation w.r.t. $\eta$. This 
is the typical action for a linear fluctuation $\overline h_m$ coupled to an external ``pump field" $\xi_m(\eta)$ (see for instance \cite{9}). 
According to this action 
the Fourier modes $u_{km}$, defined by $\nabla^2 u_{km}=-k^2u_{km}$, satisfy the
canonical evolution equation
\beq
u_{km}^{\prime \prime} + \left(k^2+m^2a^2 - {\xi_m^{\prime \prime}\over \xi_m}\right) u_{km}=0\,.
\label{32}
\eeq

Since massless and massive modes do not mix, they can be discussed  separately.
Let us first consider the massless  tensor fluctuations corresponding to the
bound state of the potential (\ref{27}), which describes long-range
gravitational interactions confined on the brane. Assuming that the perturbed
background is $Z_2$-symmetric \cite{8}, the square-integrable, $Z_2$-even
solution of Eq.~(\ref{27}) with $m=0$ is
\beq
\tpsi_0= c_0f^{p/2}\,, ~~~~~~~~~~~~~
\psi_0= c_0/ \sqrt M = {\rm const}\,,
\label{33}
\eeq 
where $c_0$ is an integration constant. By imposing that $\tpsi_0 \in ~ L^2(R)$, according to the normalization (\ref{28}), we obtain 
\beq
c_0^2= \left (\int dz f^p \right)^{-1} = (p-1)/2 z_0, ~~~~~~~~~~~
\psi_0 = \left(p-1 \over 2 M z_0 \right)^{1/2}.
\label{34}
\eeq
The effective coupling parameter for the massless mode $\overline h_0$ is determined by the action (\ref{29}), and has to be identified with the Planck mass ($\Mp \simeq 10^{19}$ GeV) controlling  long range gravitational interactions on the brane,  
\beq
 \Mp^{d-1}={M^{d-1} \over| \psi_0|^2} = {2z_0\over (p-1)} M^{d}.
\label{35}
\eeq
For $n=0$, $d=p=3$, one recovers the standard Randall-Sundrum relation \cite{4} $\Mp^2 =z_0 M^3$ between the Planck mass $\Mp$ and the bulk mass $M$. 

In order to discuss the formation of a relic background of gravitational radiation, 
amplified from the vacuum by the cosmological evolution of the brane,  we consider  a simple dynamical model where the brane geometry evolves  from an initial
(inflationary) Kasner-like regime to a final regime characterized by the
flat Minkowski metric. (A more realistic picture would require, of course, the presence of matter fields on the brane). By assuming a transition epoch localized around the
conformal time scale $\eta=-\eta_1$, the computation of the pump field
(\ref{30}) gives, for $m=0$,  
\bea
&&
\xi_0(\eta) = \left(\Mp^{d-1}\over 2 \right)^{1/2}
\left(-{\eta\over \eta_1}\right)^{1/2}\,, ~~~~~~~~~~~~~~~
\eta \ll -\eta_1\,, \nn \\
&&
\xi_0(\eta) = \left(\Mp^{d-1}\over 2 \right)^{1/2}= {\rm constant}\,, 
 ~~~~~~~~~~~~~~~ \eta \gg -\eta_1\,.
\label{36}
\eea
The effective potential of Eq.~(\ref{32})  evolves from the initial value $\xi^{\prime \prime}/ \xi=-1/(4 \eta^2)$ at $\eta \ra - \infty$ to the final
value $\xi^{\prime \prime}/ \xi=0$ at $ \eta \ra + \infty$. Note that the initial pump field, $\xi_0 
\sim (-\eta)^{1/2}$, is represented by a function of time which is independent from $d$ and $n$: this   ``universality" is typical of the Kasner solution and characterizes also  the ``minimal" pre-big bang configurations of string cosmology \cite{10}. 

The evolution of the pump field and the effective potential imply, according to Eq. (\ref{32}),  that
the Fourier amplitude of the tensor perturbation $\overline h_{k0}= u_{k0}/\xi_0$ grows logarithmically in time outside the horizon
\cite{11,12}, and that the final power of the spectrum is fixed by  the number of spatial
dimensions in which the perturbations propagate (modulo  logarithmic corrections). In fact, the massless solution (\ref{34}) for $\psi_0$ is dimensionless, so that $u_0$ has the correct canonical dimensions ($[u_0]= \Mp^{(d-1)/2}$) and can be normalized, asymptotically, to an initial spectrum of quantum vacuum fluctuations by imposing 
$u_{k0} = \exp(-i k \eta) / \sqrt{2k}$ at $\eta \ra -\infty$. Matching the 
exact Hankel solution $u_{k0}=( \pi |\eta|/4)^{1/2} H_0^{(2)} (k \eta)$ for $\eta \leq -\eta_1$ to the plane-wave solution for $\eta \geq -\eta_1$, we obtain $\overline h_{k0}= u_{k0}/\xi_0$ after the transition. The final spectral distribution is:  
\beq
\left| \da_0(k) \right|^2= k^d \left|h_{k0}\right|^2 \simeq
\left( H_1 \over \Mp\right)^{d-1} \left( k\over k_1\right)^d
\left(\ln k\eta_1\right)^2\,,
~~~~~~~
k< k_1= \eta_1^{-1}\,.
\label{37}
\eeq
(See e.g. \cite{11} for a higher-dimensional computation, and 
\cite{13} for a recent computation with the same pump field in a different background.)  

In the above equation, $H_1=k_1/a= (a \eta_1)^{-1}$ is the
curvature scale of the brane at the transition epoch $\eta_1$, and  $k_1= 
\eta_1^{-1}$ is the ultraviolet cut-off scale of the spectrum: 
high-frequency modes with $k>k_1$ are not significantly amplified by the background
transition.  The final amplitude of the relic background of massless tensor perturbations is determined by the inflation scale in Planck units. There is no difference
between this model of brane-world inflation and more conventional (Kaluza-Klein) models of inflation with
higher-dimensional factorized geometry; in particular, there is no extra
enhancement of the amplitude when the brane curvature $H_1$ exceeds the
critical value $M^d/\Mp^{d-1}$, unlike in de Sitter models of brane inflation \cite{5,6}. 

Let us now consider massive tensor fluctuations that are not localized on the
brane ($m \neq 0$). After imposing the appropriate boundary conditions at $z=0$
\cite{4,6b,7,9b} (see also Ref.~\cite{17a}), the even $Z_2$-symmetric solution
of Eq.~(\ref{27}) with $m \neq 0$ can be written in terms of first and second
kind Bessel functions as
\beq
\tpsi_m= c_m f^{-1/2} 
\left[ Y_{p-1\over2}(mz_0) J_{p+1\over2}\left(mz_0\over f\right)- 
J_{p-1\over2}(mz_0) Y_{p+1\over2}\left(mz_0\over f\right)  \right]\,.
\label{38}
\eeq
(This is a generalization to $p$ dimensions of the $3$-brane
solution presented in Ref.~\cite{7}.) The orthonormality condition 
(\ref{28}) gives \cite{7,9b,17a}
\beq
c_m= \left(mz_0\over 2\right)^{1/2} \left[Y^2_{{p-1\over 2}}(mz_0)+
J^2_{{p-1\over 2}}(mz_0)\right]^{-1/2}.
\label{39}
\eeq
It follows 
\beq
\psi_m(0)= {\tpsi_m(0)\over \sqrt M}= {F(mz_0)\over \sqrt M}, 
\label{40}
\eeq
where
\beq
F(x)= \left(x\over 2\right)^{1/2} {Y_{p-1\over2}(x) J_{p+1\over2}(x)- 
J_{p-1\over2}(x) Y_{p+1\over2}\left(x\right)\over
\left[Y^2_{{p-1\over 2}}(x)+
J^2_{{p-1\over 2}}(x)\right]^{1/2}}, ~~~~~ x=mz_0. 
\label{41}
\eeq
According to Eq. (\ref{30}), the pump field for massive modes
\beq
\xi_m(\eta)= \left(M\over 2\right)^{(d-1)/2}{\sqrt M\over F(mz_0)} 
\left(-{\eta\over \eta_1}\right)^{1/2},
\label{42}
\eeq
has different normalization, but identical time-dependence of the massless
pump field. The different coupling parameter, $M_m$, for any infinitesimal mass
interval $dm$, is defined by the action (\ref{29}) as
\beq
M_m^d= {M^{d-1} \over |\psi_m(0)|^2} = {M^d \over F^2(mz_0)}.
\label{43}
\eeq
It may be noted, incidentally, that in the light mass regime ($mz_0 \ll1$) the use of the small argument limit of the Bessel functions leads to $F^2 \simeq \Ga^{-2} [(p-1)/2] (mz_0/2)^{p-2}$, and to the effective (differential) coupling strength
\beq
8 \pi G_m dm \equiv {dm\over M_m^d} \simeq  {dm\over M^d} ~\Ga^{-2}\left(p-1\over 2\right)\left(mz_0\over 2\right)^{p-2}. 
\label{44}
\eeq
For $n=0$, $d=p=3$, one  recovers the  integral measure controlling the well known Yukawa contribution of light massive modes to short-range gravitational interactions \cite{4,7,9b},
\beq
8 \pi G_m dm \simeq {mz_0^2 \over 2 z_0M^3} dm =
8 \pi G_N {mz_0^2\over 2}  dm ,
\label{45}
\eeq
where $8 \pi G_N= \Mp^{-2}=(z_0M^3)^{-1}$ is the Newton constant on the brane. 

The cosmological amplification of the massive components can be discussed by
considering again the transition between the Kasner and Minkowski regimes at
the scale $\eta_1$. The canonical  equation (\ref{32}) is still valid, and the
evolution in time of the pump field is the same as before. If the mass term of
Eq. (\ref{32}) is negligible w.r.t.\ the transition curvature scale ($m/H_1
\sim ma_1\eta_1 \ll1$), the frequency distribution of the relativistic massive
spectrum (i.e.\ the modes with $m \ll k/a <H_1$) will be the same as in the massless case. However, the final
amplitude will generally be different, because of the different coupling
strength of massive modes. 

Let us compute the relativistic, massive mode contribution to the Fourier component $\overline h_k$ of the metric fluctuation on the brane:
\beq
\overline h_k \equiv \int^{H_1} dm ~\overline h_{km} =
 \int^{H_1} dm {u_{km} \over \xi_m}\,.
\label{46}
\eeq
(We are considering the range of modes $m \ll k/a < H_1$.) Taking into account
the definitions of $\psi_m$ and $\xi_m$, the initial normalization to a quantum
spectrum of vacuum fluctuations now imposes the condition $u_{km}=
\exp(-ik\eta)/\sqrt{kM}$ at $\eta \ra -\infty$. Performing the matching
procedure with the massive pump field (\ref{42}), we can easily obtain the
following spectral amplitude of tensor perturbations, in the differential mass
interval $dm$:

\beq
|\overline h_{km}| \simeq {F(m z_0)\over M} \left(k\eta_1\over 
kM^{d-1}\right)^{1/2} \ln(k\eta_1).
\label{47}
\eeq
Integration over $m$ leads to the final spectrum
\beq
\left| \da_m(k) \right|^2= k^d \left|\overline h_{k}\right|^2
= k^d \left(\int^{H_1} dm ~|\overline h_{km}|\right)^2  \simeq
\left( H_1 \over M_*\right)^{d-1} \left( k\over k_1\right)^d
\left(\ln k\eta_1\right)^2\,,
\label{48}
\eeq
where 
\beq
M_*^{d-1}= {M^{d-1}\over \left[\int^{H_1} {dm \over M} F(mz_0) \right]^2}=
\Mp^{d-1} {(p-1) Mz_0\over 2 \left[\int^{H_1z_0} {dx} F(x) \right]^2}\,.
\label{49}
\eeq
(We have used Eq.~(\ref{35}) for a direct comparison of $M_*$ to the Planck
mass $\Mp$.) Thus we obtain the distribution (\ref{37}), where $H_1/\Mp$ has
been replaced by $H_1/M_*$ in the renormalized amplitude. 

To discuss the importance of the relativistic massive contributions, let us
first consider the case $H_1z_0 \laq 1$, in which the integral  $I= \int dx
F(x)$ is extended to a mass range which only includes light modes with $mz_0
\leq H_1z_0 \laq 1$. In this case the integration gives a small numerical
factor, $I \laq 1$. Recalling  that  $Mz_0\sim (\Mp/M)^{p-1}$, and assuming
that our model is characterized by a strong bulk gravity with  $M \ll \Mp$  (as
required, in more general brane-world scenarios,  for a possible solution of
the hierachy problem \cite{4,9b}), we obtain $M_* \gg \Mp$. This means that the
massive contribution to the metric fluctuation spectrum is highly suppressed
w.r.t.\ the massless spectrum, in agreement with the well known decoupling of
light massive modes at low energies \cite{9b}. 

By contrast, if the cosmological transition occurs at  high curvature scale,
$H_1 z_0 \gg1$, the integral in Eq. (\ref{49}) also includes the contribution
of massive modes with $1 \ll mz_0 <H_1 z_0$. These modes  are ``light" w.r.t.\
brane curvature scale $H_1$, but ``heavy" w.r.t.\ bulk curvature scale
$z_0^{-1}$. In this case, using the properties of the Bessel functions (or
performing numerical integrations) we find that the integral in Eq. (\ref{49})
is approximated by  $\int^{H_1z_0} dx F(x) \simeq H_1 z_0/ \sqrt \pi$.  It
follows
\beq
M_*^{d-1} \sim \Mp^{d-1}~ {Mz_0 \over H_1^2 z_0^2}.
\label{50}
\eeq
In particular, $M_* <\Mp$ when 
\beq
 {Mz_0 \over H_1^2 z_0^2} \sim \left(M\over \Mp \right)^{d-1}
\left(M\over H_1 \right)^{2}<1\,.
\label{51}
\eeq
If $M<\Mp$, this condition can be satisfied for high enough transition scales
(e.g., for $H_1 \sim \Mp$). Therefore,  brane-world models with sufficiently
strong bulk gravity (i.e., sufficiently small parameter $M$) may be
characterized by strongly enhanced amplifications of metric perturbations due
to the contribution of  massive gravitons. 

A detailed study of the massive mode contribution to the perturbation spectrum, including the non-relativistic sector and the large-$m$ regime ($m>H_1$), will be
reported in a forthcoming paper \cite{16}. Here we conclude by noticing that such an enhanced amplification of tensor modes could
make the brane-world scenario considered above unstable. The amplified perturbations could indeed 
destroy the  homogeneity of the background,  and drive the brane towards a curvature singularity, unless the model is complemented by some appropriate mechanism
stopping inflation at a small enough curvature scale $H_1 \laq M_*$  (in principle even much smaller than the usual quantum gravity scale $\Mp$). This kind of instability could be relevant to brane models of pre-big bang inflation \cite{17}, where the
brane has a Kasner-like geometrical structure similar to that discussed in this paper. 

\section*{Acknowledgements} It is  a pleasure to thank V. Bozza, V. A. Rubakov and G. Veneziano for helpful discussions.


\vspace{1 cm}

\begin{thebibliography}{999}
\newcommand{\bb}{\bibitem}

\bb{1}L.P.~Grishchuk and M.~Solokin, Phys.~Rev.~D {\bf 43}, 2566 (1991). 

\bb{2}D.~Langlois, Prog.~Theor.~Phys.~Suppl.~{\bf 148}, 181 (2003). 

\bb{3}P.~Ho\v{r}ava and E.~Witten, Nucl.~Phys.~B {\bf 460}, 506 (1996); 
 Nucl.~Phys.~B {\bf 475}, 96 (1996). 

\bb{4} L.~Randall and R.~Sundrum,  Phys.~Rev.~Lett.~{\bf 83}, 4690 (1999).

\bb{5}D.~Langlois, R.~Maartens and D.~Wands, Phys.~Lett.~B {\bf 489}, 259
(2000). 

\bb{6}T.~Kobayashi, H.~Kudoh and T.~Tanaka, Phys.~Rev.~D {\bf 68}, 044025 (2003); A.~Frolov and L.~Kofman, hep-th/0209133. 

\bb{6b}R.~Easther, D.~Langlois, R.~Maartens and D.~Wands, JCAP {\bf 0310}, 014
(2003). 

\bb{7a}M. Gasperini, N. Sanchez and G. Veneziano, Nucl. Phys. B {\bf 364}, 365 (1991); Int. J. Mod. Phys. {\bf A6}, 3853 (1991).  

\bb{8a}A. V. Frolov, Phys.~Lett.~B {\bf 414}, 213 (2001); A. Fabbri, D. Langlois, D. A. Steer and R. Zegers, hep-th/0407262. 

\bb{7}V.~Bozza, M.~Gasperini and G.~Veneziano, Nucl.~Phys.~B {\bf 619}, 191
(2001). 

\bb{9a}C. Csaki, J. Erlich, T. J. Hollowood and T. Shirman, Nucl.~Phys.~B {\bf 581}, 309 (2000). 


\bb{9b}V. A. Rubakov,  Phys. Usp. {\bf 44}, 871 (2001).

\bb{8}C.~van de Bruck, M.~Dorca, R.H.~Brandenberger and A.~Lukas, Phys. Rev. D
{\bf 62}, 123515 (2000).

\bibitem{9}V.F.~Mukhanov, H.A.~Feldman and R.H.~Brandenberger, Phys.~Rep.~{\bf
215}, 203 (1992).

\bibitem{10}M.~Gasperini and G.~Veneziano, Astropart.~Phys.~{\bf 1}, 317
(1993); M.~Gasperini and G.~Veneziano, Phys.~Rep.~{\bf 373}, 1 (2003).

\bibitem{11}R.~Brustein, M.~Gasperini, M.~Giovannini, V.~Mukhanov and
G.~Veneziano, Phys.~Rev.~D {\bf 51}, 6744 (1995).

\bibitem{12}R.~Brustein, M.~Gasperini, M.~Giovannini and G.~Veneziano, Phys.~
Lett.~B {\bf 361}, 45 (1995).

\bb{13}M.~Gasperini, M.~Giovannini and G.~Veneziano, Nucl.~Phys.~B {\bf 694}, 206 (2004). 

\bb{17a}R. A. Battye, C. van der Bruck and A. Mennim, 
Phys. Rev. D {\bf 69}, 064040 (2004).

\bb{16}M.~Cavagli\`a, G.~De Risi and M.~Gasperini, in preparation. 

\bb{17}S. Foffa, Phys. Rev. D {\bf 66}, 063512 (2002); Phys. Rev. D {\bf 68},  043511 (2003). 

\end{thebibliography}
\end{document}